\documentclass[journal,10pt,onecolumn]{IEEEtran}
\IEEEoverridecommandlockouts

\usepackage{cite}
\usepackage{amsmath,amssymb,amsfonts,float,graphicx,steinmetz,bm,subfigure,multicol,multirow,physics,mathtools,multicol,multirow,url,color,xcolor,etoolbox,ifthen,cite,hyperref}
\usepackage{textcomp}
\usepackage{tikz}

\newcommand{\bi}{\begin{itemize}}
\newcommand{\ei}{\end{itemize}}

\newcommand{\be}{\begin{enumerate}}
\newcommand{\ee}{\end{enumerate}}
\newcommand{\bd}{\begin{description}}
\newcommand{\ed}{\end{description}}
\newcommand{\bc}{\begin{center}}
\newcommand{\ec}{\end{center}}
\newcommand{\bt}{\begin{tabbing}}
\newcommand{\et}{\end{tabbing}}
\newcommand{\bfig}{\begin{figure}}
\newcommand{\efig}{\end{figure}}
\newcommand{\beq}{\begin{equation}}
\newcommand{\beqarr}{\begin{eqnarray}}
\newcommand{\beqarrn}{\begin{eqnarray*}}
\newcommand{\eeq}{\end{equation}}
\newcommand{\eeqarr}{\end{eqnarray}}
\newcommand{\eeqarrn}{\end{eqnarray*}}
\newcommand{\bflr}{\begin{flushright}\vspace{-0.2in}}
\newcommand{\eflr}{\end{flushright}}
\newcommand{\bsub}{\begin{subequations}}
\newcommand{\esub}{\end{subequations}}
\newcommand{\barr}{\begin{array}}
\newcommand{\earr}{\end{array}}

\definecolor{lime}{HTML}{A6CE39}
\DeclareRobustCommand{\orcidicon}{%
    \begin{tikzpicture}
    \draw[lime, fill=lime] (0,0)
    circle [radius=0.16]
    node[white] {{\fontfamily{qag}\selectfont \tiny ID}};
    \draw[white, fill=white] (-0.0625,0.095)
    circle [radius=0.007];
    \end{tikzpicture}
    \hspace{-2mm}
}

\foreach \x in {A, ..., Z}{%
    \expandafter\xdef\csname orcid\x\endcsname{\noexpand\href{https://orcid.org/\csname orcidauthor\x\endcsname}{\noexpand\orcidicon}}
}


\begin{document}

\title{Single-photon-memory measurement-device-independent quantum secure direct communication}

\author{Xiang-Jie Li\orcidA{}, Dong Pan\orcidB{}, Gui-Lu Long\orcidC{}, and Lajos Hanzo\orcidD{},
\thanks{This work is supported by the National Natural Science Foundation of China under Grants No. 11974205 and No. 12205011, the Key R\&D Program of Guangdong province (2018B030325002), Beijing Advanced Innovation Center for Future Chip (ICFC), Tsinghua University Initiative Scientific Research Program. }
\thanks{Xiang-Jie Li is with State Key Laboratory of Low-dimensional Quantum Physics and Department of Physics, Tsinghua University, Beijing 100084, China.}

\thanks{Dong Pan is with Beijing Academy of Quantum Information Sciences, Beijing 100193, China.}

\thanks{Gui-Lu Long is with State Key Laboratory of Low-dimensional Quantum Physics and Department of Physics, Tsinghua University, Beijing 100084, China; Beijing Academy of Quantum Information Sciences, Beijing 100193, China; Frontier Science Center for Quantum Information, Beijing 100084, China; Beijing National Research Center for Information Science and Technology, Beijing 100084, China. (e-mail:gllong@mail.tsinghua.edu.cn).}

\thanks{Lajos Hanzo is with School of Electronics and Computer Science, University of Southampton, Southampton SO17 1BJ, United Kingdom.}

}

\maketitle

\begin{abstract}
   Quantum secure direct communication (QSDC) uses the quantum channel to transmit information reliably and securely. In order to eliminate the security loopholes resulting from practical detectors, the measurement-device-independent (MDI) QSDC protocol has been proposed. However, block-based transmission of quantum states is utilized in MDI-QSDC, which requires practical quantum memory that is still unavailable at the time of writing. For circumventing this impediment, we propose a single-photon-memory MDI QSDC protocol (SPMQC) for dispensing with high-performance quantum memory. The performance of the proposed protocol is characterized by simulations considering realistic experimental parameters, and the results show that it is feasible to implement SPMQC by relying on present-day technology.
\end{abstract}

\IEEEpeerreviewmaketitle

\section{Introduction}

\IEEEPARstart{Q}{uantum} communication relies on quantum physical principles to ensure secure transmission of information. Bennett and Brassard proposed the first quantum key distribution (QKD) protocol in 1984~\cite{BB84}, which is deemed to be the earliest form of quantum communication conceived for secure key negotiation between two distant legitimate users. An important step in secure communication, namely the negotiation of a symmetric random secret key, is constituted by QKD. The secret key is then used for the one-time-pad encryption of messages to obtain the ciphertexts, but the transmission of  ciphertext relies on classical communication.
  
Quantum secure direct communication (QSDC) is another quantum communication protocol that uses a quantum channel for directly transmitting secret messages. But indeed, QSDC can also be utilized for distributing secret keys from the transmitter to the receiver. In 2000, Long and Liu proposed the first QSDC protocol ~\cite{LL00}, in which the messages are mapped onto block-based entangled pairs. Then, in 2003, Deng \emph{et al.}~\cite{TwoStep03} developed the so-called two-step QSDC protocol for encoding the messages. Inspired by these protocols, many other entanglement-based QSDC protocols have been conceived, such as the high-dimensional QSDC protocol of ~\cite{wang2005high}, the multi-step QSDC protocol of ~\cite{wang2005multi}, the quantum secure direct dialogue protocol of ~\cite{twoQSDD}, the single-photon-memory two-step QSDC of ~\cite{pan2020single}, the continuous-variable QSDC of ~\cite{cao2021continuous}, the QSDC network of ~\cite{qi202115}, and so on. A QSDC protocol based on a single photon was proposed by Deng and Long in 2004 ~\cite{DL04}, which is termed as the DL04 protocol. A review of QSDC protocols was given in Ref.~\cite{long2007quantum}.

Recently, there have been several experimental demonstrations of QSDC both in fiber ~\cite{hu16experimental,zhu2017experimental,qi2019} and in free-space~\cite{pan2020experimental,zhang17experimental} relying on the QSDC protocols proposed in Refs.~\cite{TwoStep03,deng2004bidirectional,DL04}. In 2021 multi-user QSDC networks relying on secure-repeaters or on direct links~\cite{long2020globecom,qi202115,long2022evolutionary} have been reported. It has been demonstrated that a secure and reliable communication can be achieved over a noisy quantum channel in the face of eavesdropping based on Wyner's wiretap theory~\cite{wyner1975wire,qi2019,wu2019security,ye2021generic,wu2021quantum}. 

But again, in the original QSDC protocols, quantum memory is required so that the level of security can be examined in the course of transmission. However, practical quantum memory~\cite{sun2018design,sun2020qmf} is still decades away. To solve this difficulty, quantum-memory-free (QMF) protocols were proposed. Explicitly, in the QMF QSDC~\cite{sun2018design,sun2020qmf}, the information is transformed by the transmitter to the ciphertext using a shared secure transmission sequence (SSTS), which is similar to a secret key. Then using module 2 additions, Alice encodes the ciphertext into a codeword, which is mapped into quantum states, which are transmitted to Bob. Both then demodulates, decodes and recovers the plaintext message. The SSTS is then extracted from the ciphertext by the pair of communicating parties for later transmission. In the communication process, the coding efficiency and the length of the SSTS extracted from the ciphertext are determined by the channel's security capacity, which can be calculated from the error rate. The QMF coding scheme enables simultaneous ciphertext transmission and SSTS negotiation. In this scheme, the ciphertext bits can be transmitted one by one upon mapping them to the quantum state, hence eliminating the requirement for block-based transmission of quantum states. The security of the message will be guaranteed by SSTS encryption, which is effectively the classic one-time-pad encryption. QMF-DL04-QSDC using dynamic joint encryption and error-control coding has been experimentally demonstrated over a maximum communication distance of 18.5 km~\cite{sun2020qmf}. QMF coding is suitable not only for single-photon QSDC, but also for entanglement-based QSDC~\cite{pan2020single}. 

There is always a gap between theory relying on idealized simplifying assumptions and practice in any technology, hence quantum communication is no exception. Hence, realistic imperfect devices cannot meet the idealized simplifying assumptions of theory, which might lead to security loopholes. The measurement-device-independent version of quantum communication protocols~\cite{lo2012measurement,braunstein2012side,zhou2020measurement,niu2018measurement,niu2020security,das2020improving,pan2020simultaneous} bridges this gap between theory and practice by removing the detector-side channels. The MDI QSDC protocol uses quantum teleportation and message encoding to send the messages ~\cite{zhou2020measurement}. It has a pair of eavesdropping detection facilities. The first one is used to detect whether there is an eavesdropper in the vicinity before encoding the information, while the second one for integrity detection, namely for detecting whether the transmitted information is tampered with. The eavesdropping detection relies on block-based transmission, since some samples of qubits will be randomly chosen for eavesdropping detection, and the remaining qubits in the block will wait for the results of checking in quantum memory. However, similar to the original QSDC protocol's block-based transmission regime, it is difficult to realize by current quantum memory technology. Normally, we use an optical delay line instead of quantum memory to store photons~\cite{zhu2017experimental}, which inevitably introduces high attenuation. Hence, for the practical application of MDI QSDC, the conception of a new quantum-memory-free coding-assisted MDI QSDC protocol is essential. 

\section{\label{sec:level2}Details of our protocol\protect\\ 
}

\begin{figure*}
   \centering
    \includegraphics[width=\textwidth]{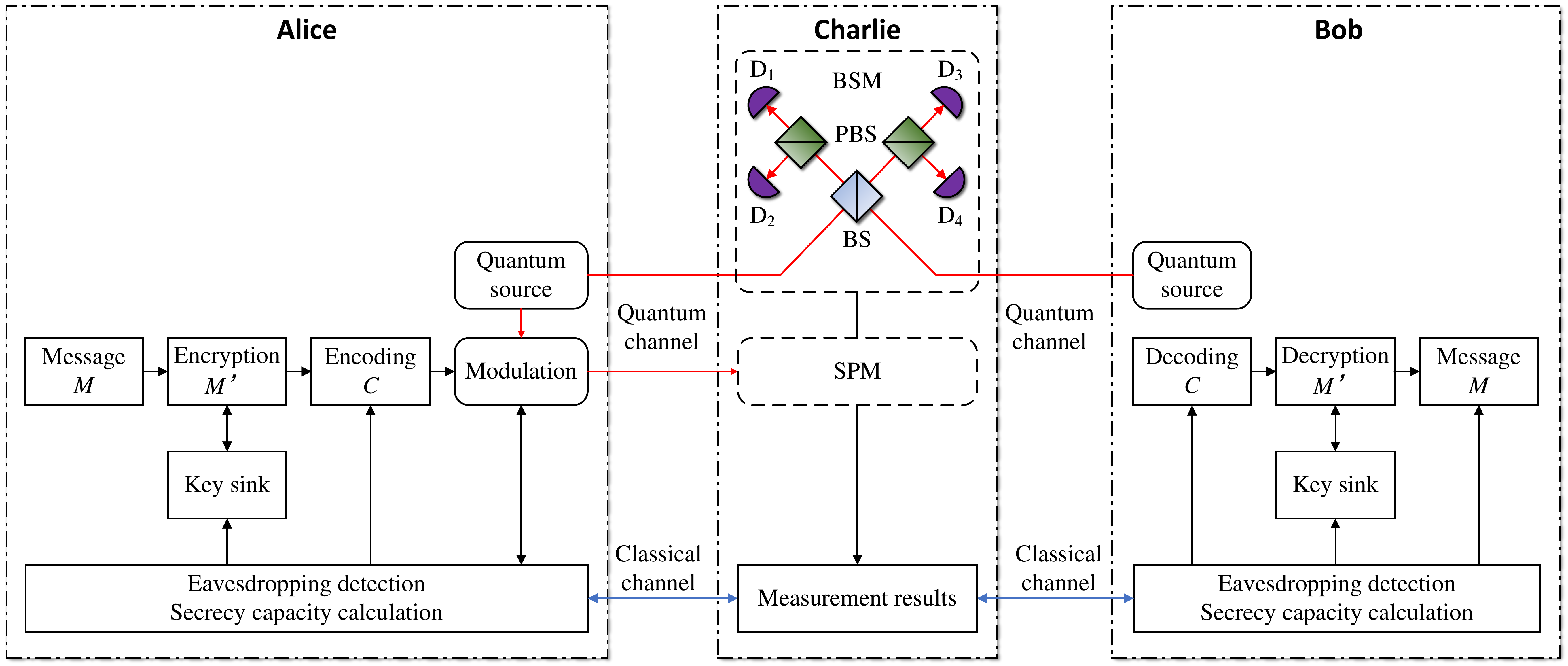}
    \caption{\label{fig:wide1}Schematic diagram of SPMQC. BS, beam splitter; BSM, Bell-state measurement; D$_{1}$, D$_{2}$, D$_{3}$, D$_{4}$, detector; PBS, polarizing beam splitter; SPM, single photon measurement.}
\end{figure*}

Here we use QMF coding to replace the quantum memory required for block-transition-based MDI QSDC. Our SPMQC-DL04 protocol is illustrated in Fig.~\ref{fig:wide1}. We use the following single qubit states:$\vert0\rangle$, $\vert1\rangle$, $\vert\pm \rangle=(\vert0\rangle\pm\vert 1\rangle)/\sqrt{2}$, and $\vert\widetilde{\pm}\rangle=(\vert0\rangle\pm i\vert1\rangle)/\sqrt{2}$. The four Bell-basis states are, $\vert\phi^\pm\rangle =(\vert 00\rangle \pm \vert 11\rangle)/\sqrt{2} $, $\vert \psi^\pm\rangle =(\vert 01\rangle\pm\vert10\rangle)/\sqrt{2} $ and $M\in\{0,1\}^m$ represents the message that Alice wants to transmit to Bob. Furthermore, $K\in\{0,1\}^m$ represents the keys in the key sink used for encryption and decryption, while $M'\in\{0,1\}^m$ and $C\in\{0,1\}^c$ represent the ciphertext and code word, respectively. We divide the code word into frames, and each round of communication includes the following six steps to send a frame of information. We note that the key sink is empty in the first round. So Alice and Bob should select the appropriate values to estimate the secrecy capacity and the key $K$~\cite{sun2020qmf}.

\textbf{Step 1, state preparation.} Alice randomly prepares a Bell-state $\vert\psi^-\rangle$ or a single photon state which is randomly in one of the four states $\left \{ \vert+\rangle_{\rm A}, \vert-\rangle_{\rm A}, \vert0\rangle_{\rm A}, \vert1\rangle_{\rm A} \right \}$. The entangled state and single photon state is used for information transmission and security check, respectively. Bob prepares a single photon state, which is randomly in one of the four states $\left \{ \vert+\rangle_{\rm B}, \vert-\rangle_{\rm B}, \vert0\rangle_{\rm B}, \vert1\rangle_{\rm B} \right \}$.

\textbf{Step 2, transmission and measurement.} Alice and Bob send their own qubit to Charlie at the same time. Note that if Alice has prepared a pair of entangled qubits, one of the qubits will be forwarded to Charlie and the remaining one is retained by Alice. Charlie receives the qubits sent by Alice as well as Bob, and then he performs a Bell-state measurement (BSM) and announces the measurement result through classical channels. 

\textbf{Step 3, security check.} If Alice prepares a single photon, she informs Bob through the classical channel and then they complete the eavesdropping detection and estimate the detection bit error rate (DBER), as detailed in Ref.~\cite{zhou2020measurement}. This procedure completes the first security check. If the DBER is below the maximum tolerable threshold, then they move to the next step. Otherwise, they return to \textbf{Step 1}. 
\begin{table}[H]
\centering
\caption{Correspondence between Alice's states and Bob's initial state as well as BSM results~\cite{zhou2020measurement}. For instance, if the BSM result is $|\psi^{+}\rangle$ and Bob's initial state is $|0\rangle_{\rm B}$, the retained qubit of Alice will be $|0\rangle_{\rm A}$. The state of qubit retained by Alice is only known to Bob, since the initial state is prepared by Bob.}
\begin{tabular}{|p{60pt}|p{25pt}|p{25pt}|p{25pt}|p{20pt}|}
\hline
\hline
Bob's initial state &$|\psi^{-}\rangle$ &$|\psi^{+}\rangle$ &$|\phi^{-}\rangle$ &$|\phi^{+}\rangle$\\
\hline

$|0\rangle_{\rm B}$ &$-|0\rangle_{\rm A}$ &$|0\rangle_{\rm A}$ &$-|1\rangle_{\rm A}$ &$-|1\rangle_{\rm A}$\\

$|1\rangle_{\rm B}$ &$-|1\rangle_{\rm A}$ &$-|1\rangle_{\rm A}$ &$-|0\rangle_{\rm A}$ &$|0\rangle_{\rm A}$\\

$|+\rangle_{\rm B}$ &$-|+\rangle_{\rm A}$ &$|-\rangle_{\rm A}$ &$-|+\rangle_{\rm A}$ &$|-\rangle_{\rm A}$\\

$|-\rangle_{\rm B}$ &$-|-\rangle_{\rm A}$ &$|+\rangle_{\rm A}$ &$|-\rangle_{\rm A}$ &$-|+\rangle_{\rm A}$\\
\hline
\hline
\end{tabular}
\label{table:measurement-results}
\end{table}
\textbf{Step 4, state recovery and transmitted message coding.} As shown in Table~\ref{table:measurement-results}, if Alice prepares an entangled state, the qubit which is retained by her will have one of the four single-photon states with equal probabilities after Charlie's BSM. Bob announces the basis that he used for prepareing the initial state. Alice then recovers the qubit she retained by performing the unitary operation $U_{T}\in\{I, i\sigma_{y}\}$ according to the BSM result. To elaborate a little further, if Bob's initial state is prepared in the basis $Z$ and the BSM result is $|\phi^{-}\rangle$ or $|\phi^{+}\rangle$, applying a unitary operation $U_{T}=i\sigma_{y}$ to Alice' qubit will transform it to the same state as Bob's initial state. We refer to this step as the state recovery, since it completes the teleportation of Bob's initial state to Alice. Alice applies the exclusive or (XOR) operation both to the message $\mathnormal{M}$ and to the key $\mathnormal{K}$ distilled from the previous round of information, for producing an encrypted ciphertext $\mathnormal{M'}$, where $\mathnormal{M'=M \oplus K}$. Next Alice performs dynamic joint encryption and error-control coding~\cite{sun2020qmf} to ensure the secure and reliable transmission of the ciphertext. Then Alice maps the ciphertext onto the qubit by using $U_{m}\in\{I, i\sigma_{y}\}$, where $\mathnormal{I}$ is used for bit 0, $i\sigma_y$ for bit 1. Alice also randomly chooses some qubits for transmission to carry out a subsequent integrity check, rather than mapping them to the ciphertext.

\textbf{Step 5, qubit transmission and measurement. } Alice sends the qubit containing the ciphertext to Charlie. Charlie measures it in the specific preparation basis that Bob has announced and publishes the measurement results through the classical channel. 

\textbf{Step 6, message decoding and integrity check.} Alice announces the random check bits and their position via the classical channel. Then the quantum bit error rate (QBER) is estimated both by Alice and Bob. 
If the QBER is below the maximum tolerable threshold, the ciphertext transmission is deemed reliable. This procedure represents the integrity check. Then Bob decodes the ciphertext to get the message $\mathnormal{M}$. If the message frame has not been transmitted, they return to \textbf{Step 1}. Otherwise, Alice and Bob use the DBER and the QBER to estimate the secrecy capacity of the current round of communication and calculate the number of keys they can distill from the ciphertext. Finally, both of them distill and insert the same keys into the key sink to encrypt and decrypt the next round of transmission. The current round of communication is over. Note that the recently proposed solution of increasing the channel capacity using masking (INCUM) ~\cite{long2021drastic} in QSDC can be invoked for improving performance as detailed in \ref{sec:INCUM}.

Note that only a single-photon storage is required for the qubit retained by Alice, which can be realized by an optical delay line. This is a common characteristic of entanglement-based protocols ~\cite{E91,BBM92,pan2020single}. In this sense, we simply term the proposed protocol as SPMQC.

\section{\label{sec:level3}Performance analysis\protect\\ }

Based on the security analysis of MDI DL04 QSDC ~\cite{niu2020security}, there exists a secrecy capacity $C_s$, which allows us to use a forward encoding scheme having a coding rate lower than $C_s$ to transmit the message reliably and securely to receivers. The associated asymptotic secrecy capacity lower bound $C_s$ is given by ~\cite{niu2020security}
    \begin{equation}
        \mathnormal{C_s}=
\mathnormal{Q}[1-h(e)-gh(\epsilon_u)],
    \end{equation}
where $Q$ is the signal gain of Bob for message decoding and $g$ is the gain difference between the channels of $AB$ and $AE$, while $e$ and $\epsilon_u$ represent the QBER and DBER, respectively. This is presented in more detail in \ref{app:Derivation details of secrecy capacity}. The DBER $\epsilon_u$ originates from the three bases $u\in\{X, Y, Z\}$ that are used for the security check~\cite{niu2020security} in \textbf{Step 3}. Subsequently, Alice chooses one of the unitary operations $U_m\in\{\sigma_x, i\sigma_y, \sigma_z\}$ for encoding bit 1 in \textbf{Step 4}, and the paired qubits containing the eigenstate of the encoding operation $\sigma_u$ will be discarded. However, this optimal procedure has not been taken into account in our protocol's description. 

\begin{figure}[htbp]
    \centering
     \includegraphics[width=8.5cm]{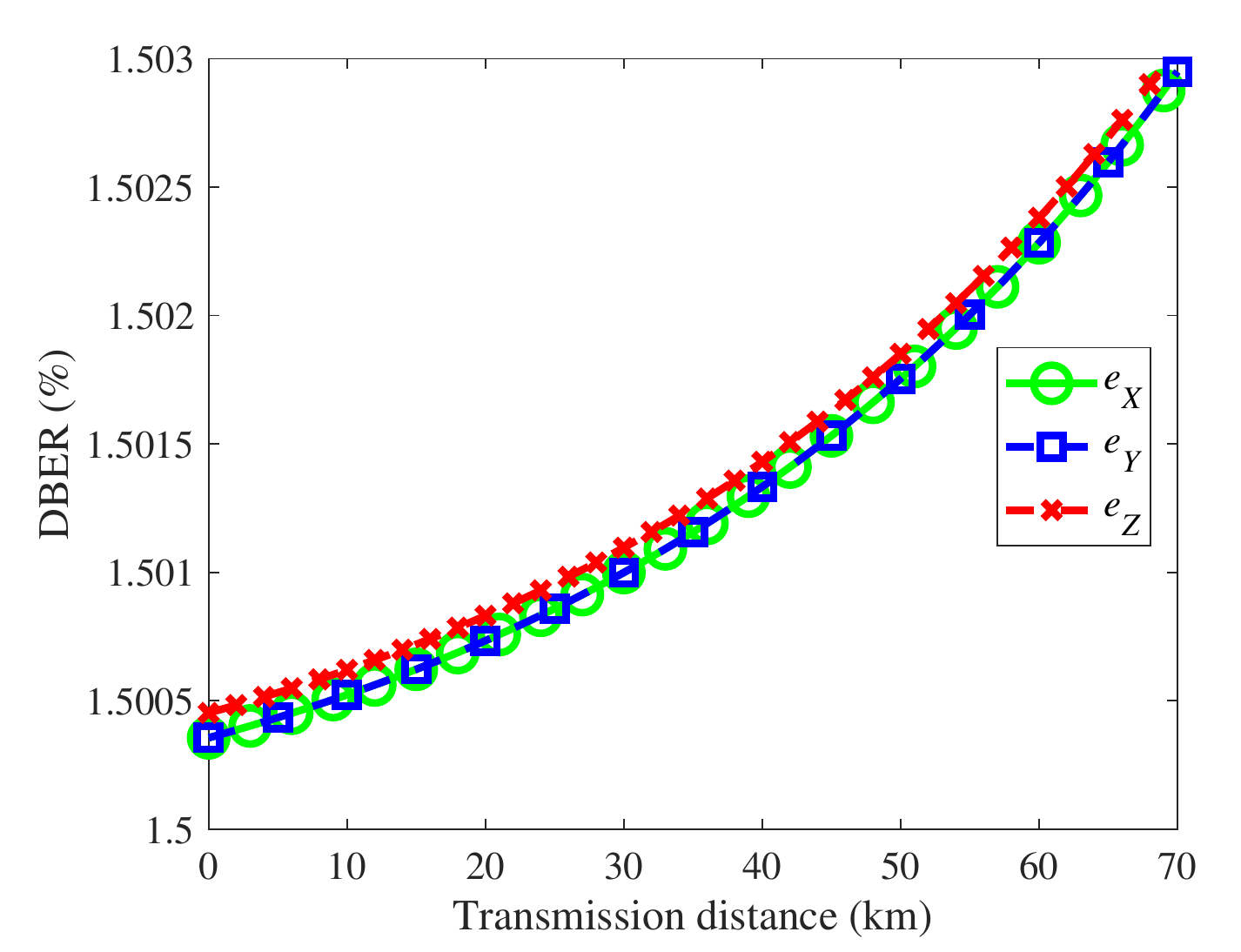}
     \caption{\label{fig:DBER}The DBER vs. the transmission distance parameterized by three different security check bases. The green line labeled by circles represents the DBER $e_X$ changing with transmission, while the blue dotted line labeled by squares and the red dotted line labeled by crosses represents $e_Y$ and $e_Z$, respectively.}
\end{figure}

\begin{figure}
   \centering
    \includegraphics[width=8.5cm]{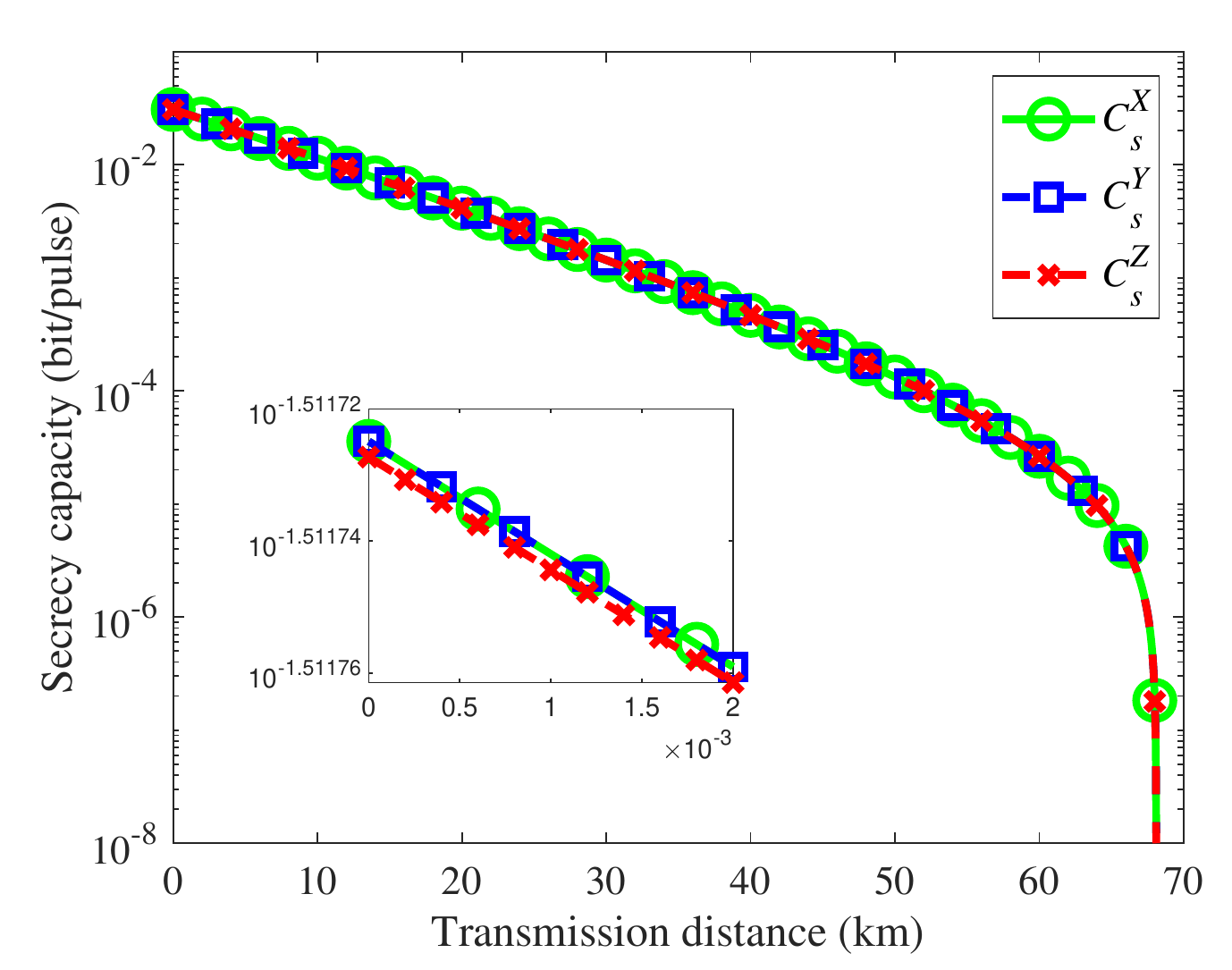}
    \caption{\label{fig:secrecy-capacity}The secrecy capacity of our SPMQC protocol vs. distance, parameterized by three different security check bases. There are three different lines. The green line labeled by circles represents the secrecy capacity $C^X_s$ changing with transmission distance, while the blue dotted line labeled by squares and the red dotted line labeled by crosses represents $C^Y_s$ and $C^Z_s$, respectively.}
\end{figure}

To determine the DBER of the proposed protocol using different bases, we perform simulation under the assumption of having ideal quantum sources. The detail derivation is presented in \ref{app:performance}. The key parameter settings for our simulations are shown in Table~\ref{Tab:parameter}.
\begin{table}[htbp]
\centering
\caption{\label{Tab:parameter}Key parameter settings of simulation.}
\begin{tabular}{ccc}
\hline 
\hline
Parameter & Value  & Description                     \\ \hline 
$\delta$   & 0.2 dB/km   & the attenuation coefficient                  \\
$\eta_d$  & 60\%    & the efficiency of detectors                      \\
$e_{0}$   & $1/2$  & the error rate of background        \\
$e_{\rm det}$  & 1.31\%       & the intrinsic detector error rate    \\
$p_d$     & $1\times10^{-6}$        & dark count   \\
$e_d$       & 0.015 & the misalignment probability         \\
\hline
\hline
\end{tabular}
\end{table}
 As seen in Fig.~\ref{fig:DBER}, the DBER $e_Z$ is higher than $e_X$ and $e_Y$. Both the DBER $e_X$ and $e_Y$ exhibit the same trends. This means that Alice and Bob will have different secrecy capacity, if they use different security check bases. The difference between $C^Z_s$ and $C^X_s$ ($C^Y_s$) can only be seen, when we focus our attention on a very short distance, as shown in the inset graph of Fig.~\ref{fig:secrecy-capacity}. Hence, we can choose an optimized security check basis $X$ or $Y$ to obtain a higher secrecy capacity. Note that the above results were obtained under the simplifying assumeption of having ideal quantum sources. It is plausible that the difference between $C^Z_s$ and $C^X_s$ ($C^Y_s$) will be larger, if practical light sources, such as weak-coherent pulses and parametric down-conversion, are considered. This is because the distribution of the number photons may play an important role in the associated DBER estimation. 

Under such circumstances, the importance of choosing an optimised basis for security check becomes plausible. The results demonstrate the feasibility of SPMQC for applications in metropolitan quantum communications with a range for a few tens of kilometers.

\begin{figure}[htbp]
  \centering
   \includegraphics[width=8.5cm]{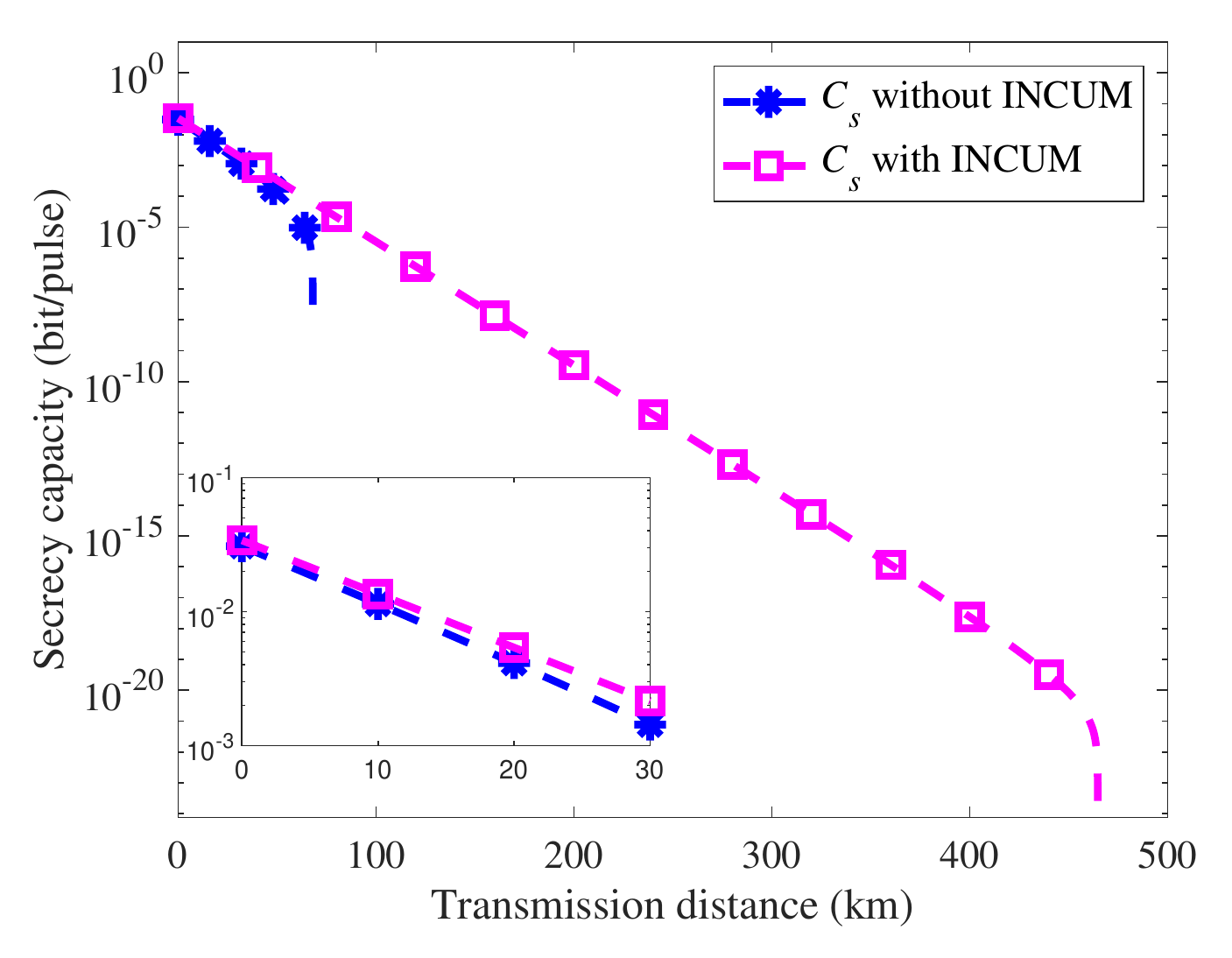}
   \caption{\label{fig:COMPARE}The secrecy capacity of our SPMQC protocol vs. distance both with and without INCUM technology. The blue line labeled by asterisks and the red line labeled by squares represent the secrecy capacity $C_s$ without and with INCUM technology changing with transmission, respectively.}
\end{figure}

\begin{figure}
   \centering
    \includegraphics[width=8.5cm]{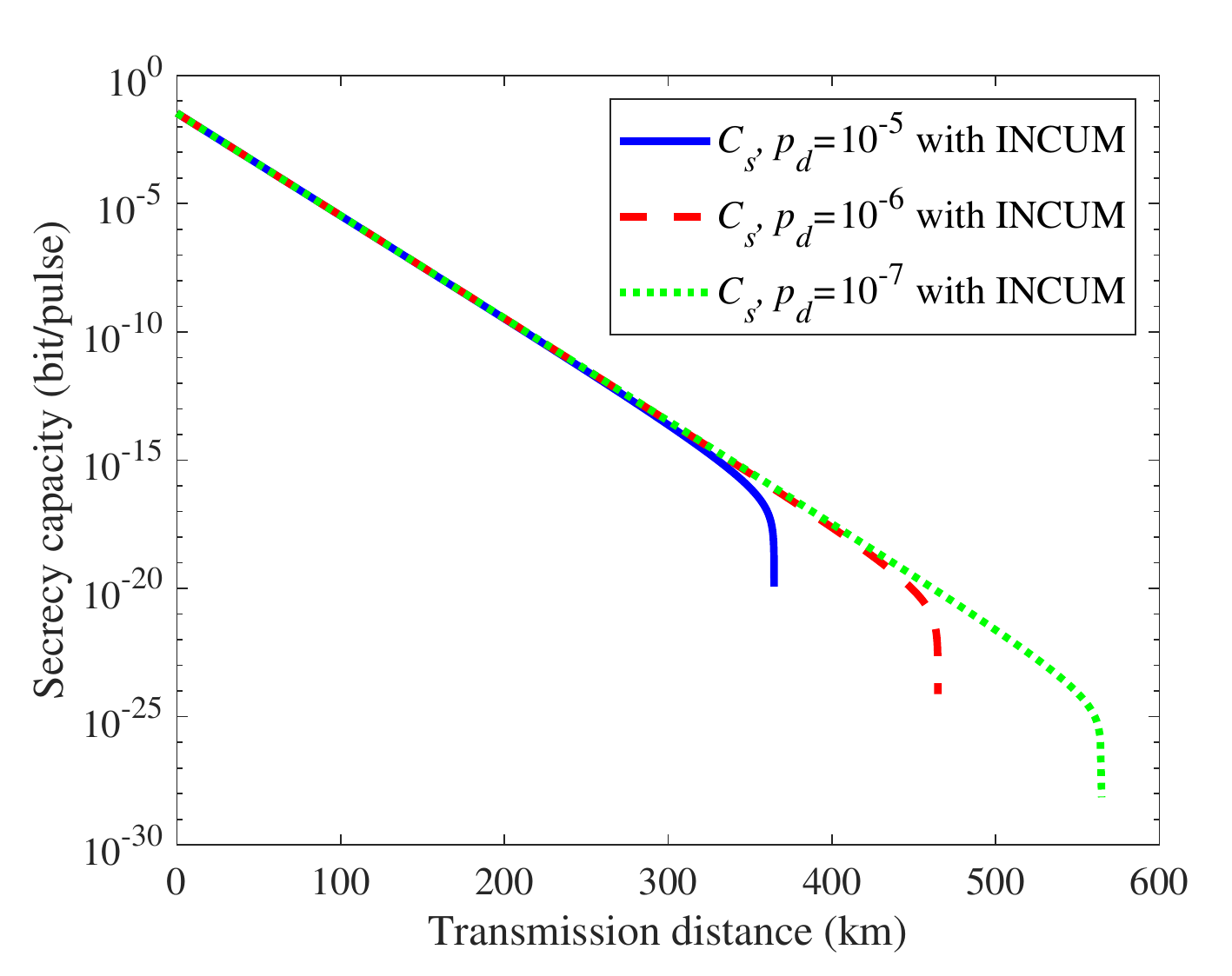}
    \caption{\label{fig:pd}The secrecy capacity of SPMQC protocol with different dark count rate. The blue line represents the secrecy capacity $C_s$ changing with transmission distance in the dark count rate of $1\times10^{-5}$, while the red and the green line  represents $1\times10^{-6}$ and $1\times10^{-7}$, respectively.}
 \end{figure}

Fig.~\ref{fig:COMPARE} shows the secrecy capacity of the SPMQC protocol both with and without increasing the capacity using masking (INCUM). Indeed, the INCUM technology ~\cite{long2021drastic} can substantially increase the secrecy capacity and the transmission distance of the protocol.

Finally, we performed numerical simulations for characterizing the influence of dark count on the upper limit of transmission distance. As shown in Fig.~\ref{fig:pd}, the dark count has a significant impact on the transmission distance. The secrecy capacity will be significantly reduced, when the signal is attenuated to be comparable to the dark count. Therefore, we can improve the transmission distance of our SPMQC protocol by reducing the dark count rate of the detector.

\section{Conclusion}

We have proposed the SPMQC protocol and analyzed its performance. Selecting the optimal security basis $X$ or $Y$ increases beneficially the security capacity of the proposed protocol. The results show our SPMQC is eminently suitable for metropolitan areas covering a range of a few tens of kilometers. with given the rapid evolution of experimental techniques, our SPMQC protocol has the potential of finding its way into practical applications. 

But before that, there are some further open issues for future research. Firstly, practical imperfect light sources have to be integrated into our proposed protocol. Secondly, the method of decoy-state based techniques could be utilized to estimate the error rate and reception rate. Thirdly, the family of optimal QMF coding techniques may be combined with optimal MDI protocols~\cite{2020High,wu2020high,zou2020measurement,gao2019long} for supporting high-rate and long-distance MDI QSDC.

\section{Appendix}
\subsection{Increasing the channel capacity using masking}
\label{sec:INCUM}
\begin{figure*}
  \centering
   \includegraphics[width=\textwidth]{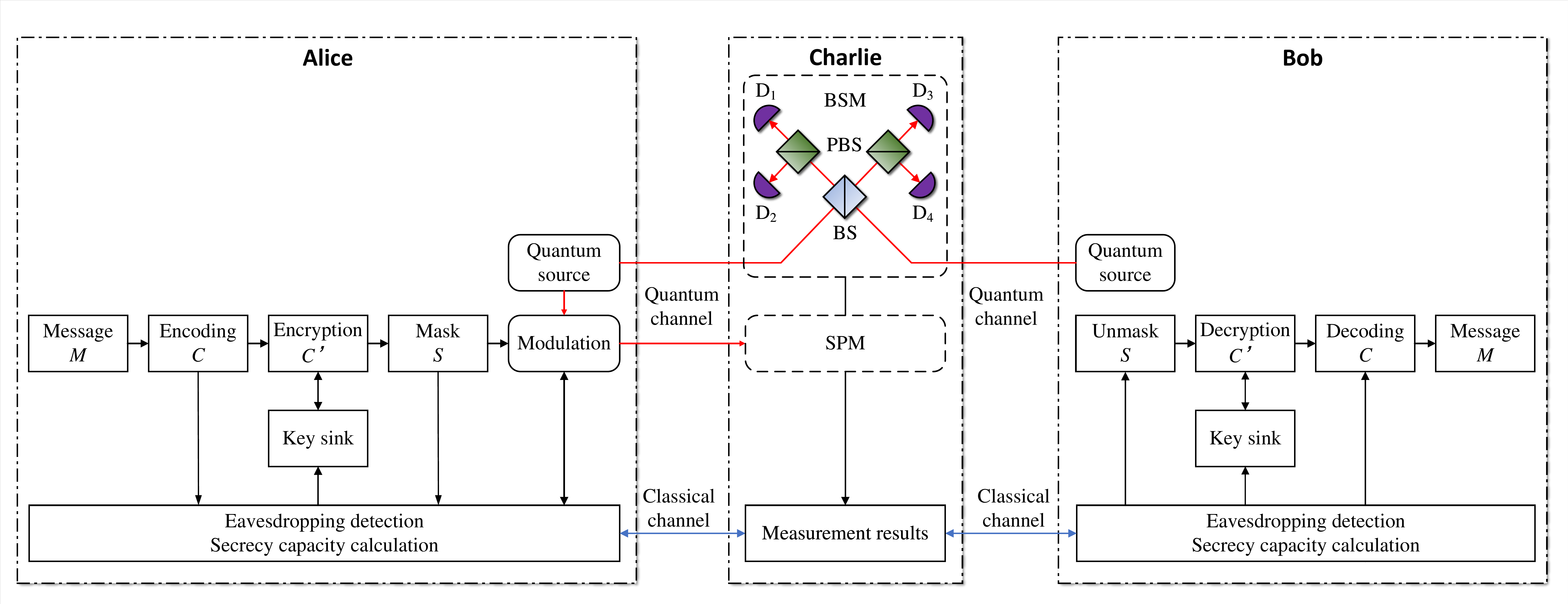}
   \caption{\label{fig:wide2}Schematic diagram of SPMQC utilize INCUM technology. $C\in\{0,1\}^c$ and $C'\in\{0,1\}^c$ represent the code word and ciphertext. $S\in\{0,1\}^c$ represents the masked message.}
\end{figure*}

Recently, Long \emph{et al.}~\cite{long2021drastic} proposed a solution to increase the channel capacity using masking (INCUM) in QSDC. We can appropriately adapt the solutions in~\cite{long2021drastic} for our encryption and encoding operation, as shown in Fig.~\ref{fig:wide2}: (1) Alice masks all data using local random numbers after the dynamic joint encryption and error-control coding process is completed in \textbf{step 4}. She generates a local random bit string $L$ and masks each qubits in the ciphertext $\mathnormal{C'}$ to form the masked message $\mathnormal{S}$, which is formulated as$S=L\oplus C'$. (2) Bob unmasks the received data after the integrity check is completed in \textbf{step 6}. Bob publicly announces the positions of his measurements, which have valid results. Based on this information, Alice publicly announces the values of the random bits mapped to the qubits in the positions announced by Bob. In such a masking scheme, Eve cannot obtain secret information for the lost photons, since the information bits mapped to lost photons are masked by local random numbers that are only known by Alice. Hence Eve can only infer secret information from the specific photons that were detected successfully by Charlie. Hence, the secrecy capacity of QSDC can be significantly improved.

\subsection{Derivation details of secrecy capacity}
\label{app:Derivation details of secrecy capacity}
In this section, we follow the analysis of ~\cite{niu2020security} to derive the secrecy capacity of the SPMQC protocol using different bases for DBER estimation and message encoding.

The maximum information Eve can infer is given by the Holevo bound~\cite{holevo1973bounds}
\begin{equation}
      I(A:E)\leq S(\sum_k p_k \rho_{AE}^k)- \sum_k p_k S(\rho_{AE}^k),
\end{equation}
where $\rho_{AE}$ is the joint state of Alice and Eve, and it is given by
\begin{equation}
      \rho_{AE}=\mathnormal{\rm Tr_B}(\vert \Psi_{ABE}\rangle \langle\Psi_{ABE}\vert)=\frac{1}{2}(P_{\vert \varphi_1\rangle}+P_{\vert \varphi_2\rangle}),
\end{equation}
where $\Psi_{ABE}$ is the system $ABE$, and $P_{\vert \varphi\rangle}$ is the projection operator of state $\vert \varphi\rangle$.

According to~\cite{niu2020security}, the definition of state $\vert \varphi_1\rangle$ and $\vert \varphi_2\rangle$ is
\begin{eqnarray}
\vert \varphi_1\rangle= \vert 0 \rangle_A (\sqrt{\delta_3}\vert E_3 \rangle+\sqrt{\delta_4}\vert E_4 \rangle)+\vert 1 \rangle_A (\sqrt{\delta_2}\vert E_2 \rangle-\sqrt{\delta_1}\vert E_1 \rangle),\nonumber\\
\vert \varphi_2\rangle=  \vert 0 \rangle_A (\sqrt{\delta_1}\vert E_1 \rangle+\sqrt{\delta_2}\vert E_2 \rangle)+\vert 1 \rangle_A (\sqrt{\delta_4}\vert E_4 \rangle-\sqrt{\delta_3}\vert E_3 \rangle).
\end{eqnarray}
After encoding, the state becomes 
\begin{eqnarray}
        \rho_{AE}^0=I\cdot \rho_{AE}\cdot I=\frac{1}{2}(\vert \varphi_1\rangle \langle\varphi_1\vert+\vert \varphi_2\rangle \langle\varphi_2\vert).\nonumber\\
        \rho_{AE}^1=\sigma_\mu\cdot \rho_{AE}\cdot \sigma^{\dagger }_\mu=\frac{1}{2}(\sigma_\mu\vert \varphi_1\rangle \langle\varphi_1\vert\sigma^{\dagger }_\mu+\sigma_\mu\vert \varphi_2\rangle \langle\varphi_2\vert\sigma_\mu^{\dagger })\equiv\frac{1}{2}(\vert \varphi_1'\rangle \langle\varphi_1'\vert+\vert \varphi_2'\rangle \langle\varphi_2'\vert).
\end{eqnarray}
If the DBER is estimated by the basis $X$ and Alice chooses $\sigma_\mu=\sigma_x$ to encode bit 1, then
\begin{eqnarray}
        \vert \varphi_1'\rangle=\vert 0 \rangle_A (\sqrt{\delta_2}\vert E_2 \rangle-\sqrt{\delta_1}\vert E_1 \rangle)+ \vert 1 \rangle_A (\sqrt{\delta_3}\vert E_3 \rangle+\sqrt{\delta_4} \vert E_4 \rangle),\nonumber\\
        \vert \varphi_2'\rangle= \vert 0 \rangle_A (\sqrt{\delta_4}\vert E_4 \rangle-\sqrt{\delta_3}\vert E_3 \rangle)+\vert 1 \rangle_A (\sqrt{\delta_1}\vert E_1 \rangle+\sqrt{\delta_2}\vert E_2 \rangle),
  \end{eqnarray}
 and 
  \begin{equation}
        \sum_k p_k \rho_{AE}^k=\frac{1}{2}\rho_{AE}^0+\frac{1}{2}\rho_{AE}^1=\frac{1}{4}(\vert \varphi_1\rangle \langle\varphi_1\vert+\vert \varphi_2\rangle \langle\varphi_2\vert+\vert \varphi_1'\rangle \langle\varphi_1'\vert+\vert \varphi_2'\rangle \langle\varphi_2'\vert).
  \end{equation}
The Gram matrix of $\sum_k p_k \rho_{AE}^k$ is~\cite{jozsa2000distinguishability}
\begin{equation}
  G_{AE}=\frac{1}{4}
\begin{bmatrix}
  1 & 0 & 0 & -\delta_1+\delta_2-\delta_3+\delta_4\\
  0 & 1 & -\delta_1+\delta_2-\delta_3+\delta_4 & 0\\
  0 & -\delta_1+\delta_2-\delta_3+\delta_4 & 1 & 0\\
  -\delta_1+\delta_2-\delta_3+\delta_4 & 0 & 0 & 1
\end{bmatrix}.
\end{equation}
By observing that $-\delta_1+\delta_2-\delta_3+\delta_4=-(1-2\epsilon_X)$~\cite{niu2020security}, the eigenvalue $\lambda_{AE}$ of $G_{AE}$ is
\begin{equation}
  \lambda _{AE}=\frac{1}{4}
\begin{bmatrix}
  1+(1-2\epsilon_X)\\
  1+(1-2\epsilon_X)\\
  1-(1-2\epsilon_X)\\
  1-(1-2\epsilon_X)
\end{bmatrix}.
\end{equation}
We then obtain the entropy 
\begin{equation}
        S\left(\sum_k p_k \rho_{AE}^k\right)=1+h(\epsilon_X).
\end{equation}
Since $\sum_k p_k S(\rho_{AE}^k)=1$, $I(A:E)$ is given by
  \begin{equation}
        I(A:E)\leq h(\epsilon_X).
  \end{equation}
Finally, the secrecy capacity becomes 
  \begin{equation}
        \mathnormal{C^X_s}\geq \mathnormal{Q}[1-\mathnormal{h(e)-gh(\epsilon_X)]}.
  \end{equation}

If the DBER is estimated by the basis $Z$ and Alice chooses $\sigma_\mu=\sigma_z$ to encode bit 1, then we have
  \begin{eqnarray}
        \vert \varphi_1'\rangle= \vert 0 \rangle_A (\sqrt{\delta_3}\vert E_3 \rangle+\sqrt{\delta_4}\vert E_4 \rangle)-\vert 1 \rangle_A (\sqrt{\delta_2}\vert E_2 \rangle-\sqrt{\delta_1}\vert E_1 \rangle)\nonumber\\
        \vert \varphi_2'\rangle=  \vert 0 \rangle_A (\sqrt{\delta_1}\vert E_1 \rangle+\sqrt{\delta_2}\vert E_2 \rangle)-\vert 1 \rangle_A (\sqrt{\delta_4}\vert E_4 \rangle-\sqrt{\delta_3}\vert E_3 \rangle),
  \end{eqnarray}
and
  \begin{equation}
        \sum_k p_k \rho_{AE}^k=\frac{1}{2}\rho_{AE}^0+\rho_{AE}^1\\
        =\frac{1}{4}(\vert \varphi_1\rangle \langle\varphi_1\vert+\vert \varphi_2\rangle \langle\varphi_2\vert+\vert \varphi_1'\rangle \langle\varphi_1'\vert+\vert \varphi_2'\rangle \langle\varphi_2'\vert).
  \end{equation}
The Gram matrix  of $\sum_k p_k \rho_{AE}^k$ is\\
\begin{equation}
  G_{AE}=\frac{1}{4}
\begin{bmatrix}
  1 & 0 & -\delta_1-\delta_2+\delta_3+\delta_4 & 0\\
  0 & 1 & 0 & \delta_1+\delta_2-\delta_3-\delta_4\\
  -\delta_1-\delta_2+\delta_3+\delta_4 & 0 & 1 & 0\\
  0 & \delta_1+\delta_2-\delta_3-\delta_4 & 0 & 1
\end{bmatrix}.
\end{equation}
We notice that $\delta_1+\delta_2-\delta_3-\delta_4=1-2\epsilon_Z$~\cite{niu2020security}. So the eigenvalue $\lambda_{AE}$ of $G_{AE}$ is
\begin{equation}
  \lambda _{AE}=\frac{1}{4}
\begin{bmatrix}
  1+(1-2\epsilon_Z)\\
  1+(1-2\epsilon_Z)\\
  1-(1-2\epsilon_Z)\\
  1-(1-2\epsilon_Z)
\end{bmatrix}.
\end{equation}
Finally, the secrecy capacity is formulated as:
\begin{equation}
        \mathnormal{C^Z_s}\geq \mathnormal{Q}[1-\mathnormal{h(e)-gh(\epsilon_Z)]}.
\end{equation}

\subsection{Performance analysis}
\label{app:performance}
To determine the secrecy capacity of the proposed protocol using different bases, we perform simulation under the assumption of having ideal quantum sources. Let us assume that $\mathnormal{\eta_c}=\eta_{d}10^{\frac{-\delta d}{10}}$ ~\cite{pan2020experimental} represents the channel transmittance, where $\eta_d$ is the detection efficiency, $\delta$ is the attenuation coefficient of the fiber, and $d$ is the transmission distance. Furthermor, $p_d$ is the dark count of single-photon detectors and we assume that $g=1/\eta_c$. If we use the method of increasing capacity using masking ~\cite{long2021drastic}, the correlation between the qubits stolen by Eve and the qubits received by Bob will be cut off. This means that the difference between the channels of $AB$ and $AE$ will disappear, i.e. $g=1$. In our protocol, Charlie performed measurement, which are part of \textbf{Step 2} and \textbf{Step 5}, respectively. Henus, the gain of Bob upon decoding the message is given by 
\begin{equation}
Q=p_A(n_A)p_B(n_B)q_{C1}q_{C2},
\end{equation}
where $p_A(n_A=1)=p_B(n_B=1)=1$ represents having ideal quantum sources, while $q_{C1}$ and $q_{C2}$ is the click rate of Charlie upon receiving photons from Alice as well as Bob \textbf{Step 2} and in \textbf{Step 5}, respectively. According to the model of ~\cite{wang2014simulating}, the gain for the different basis of $u\in\{X, Y, Z\}$ is given by 
\begin{eqnarray}
        G_u=\frac{1}{4}\sum_{(i, j)\in \mathcal{S}}\left(\mathcal{P}_{ij}^{\zeta_1\zeta_2 }+\mathcal{P}_{ij}^{\zeta_2\zeta_1 }+\mathcal{P}_{ij}^{\zeta_1\zeta_1 }+\mathcal{P}_{ij}^{\zeta_2\zeta_2 }\right),
\end{eqnarray}
where $(i, j)$ is the detector index in Fig.1. It represents the successful events of Bell-state measurements containing the two-fold click of detectors and $\mathcal{S}=\{(1, 4), (2, 3), (1, 2), (3, 4)\}$, while $\zeta_1$ and $\zeta_2$ denote the incident polarization eigenstates of basis $u$. $\zeta_1$ and $\zeta_2$ could be $H$, $V$, $+$, $-$, $\widetilde{+}$ or $\widetilde{-}$, where $H$ and $V$ represent the marks of horizontal and vertical states, while $|+\rangle=(|H\rangle+|V\rangle)/\sqrt{2} $, $|-\rangle=(|H\rangle-|V\rangle)/\sqrt{2}$, $|\widetilde{+}\rangle=(|H\rangle+i|V\rangle)/\sqrt{2}$, and $|-\rangle=(|H\rangle-i|V\rangle)/\sqrt{2}$. To elaborate a little further, for a successful Bell-state measurement the detector $D_1$ and $D_4$ or $D_2$ and $D_3$ of Charlie will successfully detect the signal if Alice and Bob respectively send out the polarized photon $+$ and $-$ or $-$ and $+$ in X basis

In the linear channel attenuation model, $\mathcal{P}_{ij}^{\zeta_1\zeta_2 }$ is given by
    \begin{eqnarray}
    \label{Eq:Pij}
        \begin{aligned}
        \mathcal{P}_{ij}^{\zeta_1\zeta_2 }=\sum_{k_1, k_2}[C^{k_1}_{n_{A}}\eta_{c}^{k_1}(1-\eta_c)^{n_{A}-k_1}\\
        \times C^{k_2}_{n_{B}}\eta_{c}^{k_2}(1-\eta_c)^{n_{B}-k_2}p_{ij}^{\zeta_1 \zeta_2 }(k_1, k_2)],
        \end{aligned}
    \end{eqnarray}
where the single-photon source of Alice and Bob are associated with the photon number of $n_A=n_B=1$, hence we have $k_1, k_1\in\{0, 1\}$, while $p_{ij}^{\zeta_1 \zeta_2 }(k_1, k_2)$ is the conditional probability of the specific event $(i, j)$. We can deduce the result of $\mathcal{P}_{ij}^{\zeta_1\zeta_2 }$ given that the value of $p_{ij}^{\zeta_1 \zeta_2 }(k_1, k_2)$ can be found in ~\cite{wang2014simulating}, see the Sec.~\ref{sec:pij} for further detail. The click rate $q_{C1}$ of Charlie is given by
\begin{equation}
q_{C1}=\begin{cases}
 \frac{\eta_c}{3}(G_Y+G_Z)& \text{ if } \epsilon_X  \text{\;is estimated} \\ 
 \frac{\eta_c}{3}(G_X+G_Z)& \text{ if } \epsilon_Y  \text{\;is estimated} \\ 
 \frac{\eta_c}{3}(G_X+G_Y)& \text{ if } \epsilon_Z  \text{\;is estimated} \\ 
\end{cases},
\end{equation}
where the denominator 3 is derived from the three bases that are used in the security check. Obviously, $G_u$ has no contribution to $q_{C2}$, since the paired qubits used for eavesdropping detection will be discarded. The click rate $q_{C2}$ of Charlie is given by
\begin{equation}
        q_{C2}=\eta_c+(1-\eta_c)p_d.
\end{equation}

The QBER and DBER is quantified by the ratio of the wrong clicks to the total number of clicks. Then we have
    \begin{equation}
        e=\frac{e_0p_d+e_{\rm edt}\eta_c}{\eta_c+(1-\eta_c)p_d},
    \end{equation}
and

    \begin{equation}
        \begin{aligned}
          \label{eq:DBER}
          \hat{\epsilon_X}&=\frac{\sum\limits_{(i, j)\in\{(1,4), (2,3)\}}(\mathcal{P}_{ij}^{++}+\mathcal{P}_{ij}^{--})+\sum\limits_{(i, j)\in\{(1,2), (3,4)\}}(\mathcal{P}_{ij}^{+- }+\mathcal{P}_{ij}^{-+ })}{4G_X},\\
          \hat{\epsilon_Y}&=\frac{\sum\limits_{(i, j)\in\{(1,4), (2,3)\}}(\mathcal{P}_{ij}^{\widetilde{+}\widetilde{+}}+\mathcal{P}_{ij}^{\widetilde{-}\widetilde{-}})+\sum\limits_{(i, j)\in\{(1,2), (3,4)\}}(\mathcal{P}_{ij}^{\widetilde{+}\widetilde{-} }+\mathcal{P}_{ij}^{\widetilde{-}\widetilde{+} })}{4G_Y},\\
          \hat{\epsilon_Z}&=\frac{\sum\limits_{(i, j)\in \mathcal{S}}\left(\mathcal{P}_{ij}^{HH}+\mathcal{P}_{ij}^{VV}\right)}{4G_Z}.
        \end{aligned}
    \end{equation}

The polarization misalignment is the major source of errors for the polarization-encoding based MDI system~\cite{xu2013practical}, thus the DBER $\hat{\epsilon_u}$ should be modified as 
\begin{equation}
\epsilon_u=e_{d}(1-2\hat{\epsilon_u})+\hat{\epsilon_u},
\end{equation}
where $e_0$ is the error rate of the background and $e_{\rm det}$ represents the intrinsic detector error rates, while $e_{d}$ is the misalignment probability.

\subsection{The derivation of $\mathcal{P}_{ij}^{\zeta_1\zeta_2 }$}
\label{sec:pij}
The derivation of $\mathcal{P}_{ij}^{\zeta_1\zeta_2 }$ under different polarization states are summarized as follows~\cite{Niu2020Research}.
\begin{equation}
\mathcal{P}_{ij}^{HV}=\mathcal{P}_{ij}^{VH}=(1-\eta_c)^2p_d^2(1-p_d)^2+(1-\eta_c)\eta_cp_d(1-p_d)^2+\frac{1}{4}\eta_c^2(1-p_d)^2.
\end{equation}
\begin{equation}
\mathcal{P}_{ij}^{HH}=\mathcal{P}_{ij}^{VV}=(1-\eta_c)^2p_d^2(1-p_d)^2+(1-\eta_c)\eta_cp_d(1-p_d)^2+\frac{1}{2}\eta_c^2p_d(1-p_d)^2.
\end{equation}
\begin{equation}
\mathcal{P}_{12}^{-+}=\mathcal{P}_{12}^{+-}=\mathcal{P}_{34}^{-+}=\mathcal{P}_{34}^{+-}=(1-\eta_c)^2p_d^2(1-p_d)^2+(1-\eta_c)\eta_cp_d(1-p_d)^2+\frac{1}{4}\eta_c^2p_d(1-p_d)^2.
\end{equation}
\begin{equation}
\mathcal{P}_{14}^{-+}=\mathcal{P}_{14}^{+-}=\mathcal{P}_{23}^{-+}=\mathcal{P}_{23}^{+-}=(1-\eta_c)^2p_d^2(1-p_d)^2+(1-\eta_c)\eta_cp_d(1-p_d)^2+\frac{1}{4}\eta_c^2(p_d+1)(1-p_d)^2.
\end{equation}
\begin{equation}
\mathcal{P}_{12}^{++}=\mathcal{P}_{34}^{++}=\mathcal{P}_{12}^{--}=\mathcal{P}_{34}^{--}=(1-\eta_c)^2p_d^2(1-p_d)^2+(1-\eta_c)\eta_cp_d(1-p_d)^2+\frac{1}{4}\eta_c^2(p_d+1)(1-p_d)^2.
\end{equation}
\begin{equation}
\mathcal{P}_{14}^{++}=\mathcal{P}_{23}^{++}=\mathcal{P}_{14}^{--}=\mathcal{P}_{23}^{--}=(1-\eta_c)^2p_d^2(1-p_d)^2+(1-\eta_c)\eta_cp_d(1-p_d)^2+\frac{1}{4}\eta_c^2p_d(1-p_d)^2.
\end{equation}
The derivation of $\mathcal{P}_{ij}^{\zeta_1\zeta_2 }$ for the polarization eigenstates of basis $Y$ follows similar lines to that of the basis $X$.


\end{document}